# Emergent magnetic field and vector potential of the toroidal magnetic hopfions

Konstantin Y. Guslienko[*]

*Depto. Polímeros y Materiales Avanzados: Física, Química y Tecnología, Universidad del País Vasco, UPV/EHU, 20018 San Sebastián, Spain*

*EHU Quantum Center, University of the Basque Country, UPV/EHU, 48940 Leioa, Spain*

*IKERBASQUE, the Basque Foundation for Science, 48009 Bilbao, Spain*

**Abstract**

Magnetic hopfions are localized magnetic solitons with non-zero 3D topological charge (Hopf index). Here I present an analytical calculation of the toroidal magnetic hopfion vector potential, emergent magnetic field, the Hopf index, and the magnetization configuration. The calculation method is based on the concept of the spinor representation of the Hopf mapping. The hopfions with arbitrary values of the azimuthal and poloidal vorticities are considered. The special role of the toroidal coordinates and their connection with the emergent vector potential gauge are demonstrated. The hopfion magnetization field is found explicitly for the arbitrary Hopf indices. It is shown that the Hopf charge density can be represented as a Jacobian of the transformation from the toroidal to the cylindrical coordinates.

# 1. Introduction

Topologically non-trivial configurations of a vector field describing an order parameter can be classified by using maps from the coordinate space ($r$) to the order parameter space. The vector order parameter for a ferromagnetic media is its net magnetization $M(r)$. The magnetization field $M(r)$ in 3D space represented by the unit field vector $m(r_\alpha) = M(r)/|M(r)|$ depends, in general, on three spatial coordinates $r_\alpha$, α =1, 2 ,3. The theory of topological charges of 1D and 2D magnetization textures is well developed and used for the classification of topological magnetic solitons such as domain walls (kinks), vortices and skyrmions [1]. Topological charges describe degrees of mappings (homotopy invariants) of 1D- ($R^1$) or 2D coordinate space ($R^2$) to the unit sphere $m^2 = 1$ in the magnetization space $S_2(m)$, *i.e.*, $R^1 \rightarrow S_2(m)$, $R^2 \rightarrow S_2(m)$. Mapping of the 3D coordinate space $R^3(r)$ to the unit sphere $m^2 = 1$ ($S_2$) is more complicated.

The explicit form of the mapping $R^3 \rightarrow S_2$ was introduced by Hopf in 1931 [2]. Later it was shown that the Hopf index (a degree of the mapping $R^3 \rightarrow S_2(m)$) can be represented as some integral of the expression composed by a continuous classical field (for instance, the magnetization field $m(r)$) and its spatial derivatives [3]. Initially, the topological three-dimensional solitons with non-zero Hopf index, named Faddeev-Hopf knots or hopfions, were considered in the classical field theories. Faddeev [4] suggested a new Lagrangian (sometime referred to as the Faddeev-Skyrme´s Lagrangian), which has stable soliton solutions for the three-component classical field with conserving a topological charge (the Hopf charge). Then, de Vega found explicitly the field configurations (closed vortices) for the unit Hopf charge in the toroidal coordinates [5].

Now such hopfions are called toroidal hopfions. Nicole [6] suggested the natural form of the Hopf mapping, which allowed him to write the soliton field configuration for the unit Hopf index. Later, motivated by the seminal paper by Faddeev and Niemi [7], several papers on the toroidal hopfions in the classical field theory were published [8-12]. The toroidal hopfions attracted interest of researchers because such solitons are stable solutions of the Faddeev-Skyrme´s Lagrangian [12]. The recent review can be found in Ref. [13]. The magnetization field $\boldsymbol{m}(\boldsymbol{r})$ approaches some constant value $\boldsymbol{m}_0$ at infinity $|\boldsymbol{r}| \to \infty$ for localized solitons. The condition $\boldsymbol{m}(\boldsymbol{r}) \to \boldsymbol{m}_0$ implies that the Hopf index, which distinguishes the different homotopy classes $\pi_3(S_2) = Z$, is an integer in infinite samples [7]. Therefore, there is a class of the toroidal hopfions in ordered media, which are described by an integer Hopf index $Q_H = 0, \pm 1, \ldots$. The Hopf index of the toroidal hopfions is a product of two winding numbers, the planar winding and the twisting of the magnetization configuration, respectively. Namely, due to the property that the magnetization field is asymptotically trivial, $\boldsymbol{m}(\boldsymbol{r}) \to \boldsymbol{m}_0$ at $|\boldsymbol{r}| \to \infty$, it is possible to compactify $R^3 \to S_3$. The hopfions as localized solitons of a classical field resemble particle-like objects.

Nowadays, the hopfions are investigated in the field of condensed matter physics (magnetic media [14], liquid crystals and colloids [15], ferroelectrics [16]) as well as photonics [17], optics [18], electromagnetism and gravitation [19], etc. There is a considerable interest in 3D inhomogeneous magnetization configurations classified by a linking number of the preimages of two distinct points in $S_2(\boldsymbol{m})$ in the 3D coordinate space ($R^3$), *i.e.*, by the non-zero Hopf index. The corresponding configurations are called ¨magnetic hopfions¨. Although, 3D localized topological magnetic solitons were introduced by Dzyaloshinskii et al. [20] long time ago, the reincarnation of interest to such magnetization textures started after the papers by Sutcliffe [21, 22] relatively recently.

The simplest magnetic toroidal hopfions with $|Q_H| = 1$ were considered in infinite ferromagnetic films [23, 24] and cylindrical dots [22, 25-27] using the unit-vector field hopfion ansatz [9]. It was shown numerically that the toroidal hopfions with $|Q_H| = 1$ can be the ground state of circular chiral nanodots [26] assuming a strong surface magnetic anisotropy. Basing on the theoretical predictions [22], the first experimental observation of the magnetic hopfions was carried out in the Ir/Co/Pt multilayer systems [28]. It was shown that except the toroidal magnetic hopfions there is another class of magnetization configurations with non-zero Hopf index – Bloch points [29]. Very recently the theoretical papers on the magnetic hopfion dynamics [30, 31] were published.

In this article, we present a direct analytical calculation of the magnetic hopfion's emergent vector potential, emergent magnetic field, and the magnetization configuration. The approach is based on the definition of the Hopf mapping $R^3 \to S_2(\boldsymbol{m})$.

## 2. Results and Discussion

We start from the general definition of the emergent electromagnetic field tensor (in the units of $\hbar/2e$) resulting from an inhomogeneous spin texture $\boldsymbol{m}(\boldsymbol{r})$ [32]

$$F_{\mu\nu}(\boldsymbol{m}) = \boldsymbol{m} \cdot (\partial_\mu \boldsymbol{m} \times \partial_\nu \boldsymbol{m}), \tag{1}$$

where $\boldsymbol{m}(\boldsymbol{r}) = \boldsymbol{M}(\boldsymbol{r})/M_s$ is the unit magnetization vector, $M_s$ is the saturation magnetization, $\partial_\mu = \partial/\partial x_\mu$ denote spatial derivatives and the indices $\mu, \nu$ corresponds to the components of 3D radius-vector $\boldsymbol{r}$ in an orthogonal coordinate system.

The field tensor is related to the emergent field vector potential $\boldsymbol{A}$ as

$$F_{\mu\nu}(\boldsymbol{m}) = \partial_\mu A_\nu(\boldsymbol{m}) - \partial_\nu A_\mu(\boldsymbol{m}) . \qquad (2)$$

The emergent magnetic field $\boldsymbol{B} = \nabla \times \boldsymbol{A}$ (gyrocoupling density) can be defined as in the standard electrodynamics

$$B_\lambda(\boldsymbol{m}) = \frac{1}{2}\varepsilon_{\lambda\mu\nu}F_{\mu\nu}(\boldsymbol{m}). \qquad (3)$$

The Hopf index is then calculated as the integral over the system volume from the dot product $\boldsymbol{A} \cdot \boldsymbol{B}$ [3-7]. A general expression of the Hopf invariant for the mapping of the spheres $f: S_3 \to S_2$ is

$$Q_H = \frac{1}{(4\pi)^2}\int dV \boldsymbol{A} \cdot \boldsymbol{B}, \qquad (4)$$

The emergent magnetic field $\boldsymbol{B}$ is unambiguously defined by Eq. (3) for any given magnetization texture $\boldsymbol{m}(\boldsymbol{r})$. However, to calculate the Hopf index we need to find the emergent field vector potential $\boldsymbol{A}$ in a proper gauge. As we show below, this is possible to implement using a definition of the Hopf mapping in a spinor representation.

The definition of the Hopf mapping of the 3D coordinate space $R^3$ (represented by the unit radius hypersphere $S_3$ in the 4D space) to the unit sphere $S_2(\boldsymbol{m})$ is [33]

$$\boldsymbol{m} = Z^+ \boldsymbol{\sigma} Z, \tag{5}$$

where $\boldsymbol{\sigma} = (\sigma_x, \sigma_y, \sigma_z)$ are the Pauli matrices and $Z = (Z_1, Z_2)^T$ is a spinor composed from the hypersphere coordinates $X_i$, $i = 1, 2, 3, 4$ satisfying the condition $\sum_{i=1}^{4} X_i^2 = 1$. The spinor components are $Z_2 = X_1 + iX_2$, $Z_1 = X_4 + iX_3$, and their normalization is $|Z_1|^2 + |Z_2|^2 = 1$.

The magnetization $\boldsymbol{m}$ components in the spinor representation are defined as

$$m_x + im_y = 2Z_2 Z_1^*, \quad m_z = |Z_1|^2 - |Z_2|^2. \tag{6}$$

Such magnetization components correspond to the complex function $w = (m_x + im_y)/(1 + m_z)$ describing the stereographic projection of the complex plane $w = Re(w) + iIm(w)$ to the surface of the unit sphere $S_2(\boldsymbol{m})$, *i.e.*, the compactifying the complex plane $w$ to $S_2$. Therefore, we can write $w = Z_2/Z_1$ in the spinor representation [2]. To express the coordinate dependence of the complex function $w$ we introduce the mapping of the real coordinate space $R^3(x, y, z)$ the hypersphere $S_3(\mathbf{X})$ (compactifying $R^3$ to $S_3$) in the form [6]:

$$X_4 = \frac{(1-r^2)}{(1+r^2)}, \quad X_i = \frac{2x_i}{(1+r^2)}, \quad r^2 = \sum_i x_i^2, \quad i = 1, 2, 3. \tag{7}$$

Now, we can re-write the Hopf mapping (5) in the explicit form $\boldsymbol{m}(\boldsymbol{r}) = Z^+(\boldsymbol{r})\boldsymbol{\sigma} Z(\boldsymbol{r})$. The hopfion magnetization for the simplest hopfion with the Hopf index $|Q_H| = 1$ [6] is described via the function $w$ (all coordinates are in units of the hopfion radius $a$ serving as a natural scale parameter)

$$w(\boldsymbol{r}) = \frac{2(x+iy)}{(1-r^2+2iz)}. \tag{8}$$

Eq. (8) is a result of the mapping of the real coordinate space $R^3(\boldsymbol{r})$ to the hypersphere $S_3(\mathbf{X})$ and, then, the hypersphere $S_3(\mathbf{X})$ to the unit sphere $S_2(\boldsymbol{m})$ using the hopfion magnetization texture $\boldsymbol{m}(\boldsymbol{r})$. The radius vector $\boldsymbol{r}$ can be described in the different orthogonal coordinate system. However, the function $w(\boldsymbol{r})$ is the simplest in the toroidal coordinates $\boldsymbol{r}(\eta, \beta, \varphi)$ [5, 8, 10, 11]. There is the connection between the cylindrical $(\rho, \varphi, z)$ (or Cartesian $(x, y, z)$) and toroidal $(\eta, \beta, \varphi)$ coordinates [34]

$$\rho = a\frac{sinh(\eta)}{\tau}, \quad z = a\frac{sin(\beta)}{\tau}, \quad \varphi = \varphi, \quad \tau = cosh(\eta) - cos(\beta), \tag{9}$$

where the toroidal parameter $\eta$ varies from 0 to $\infty$, the poloidal angle $\beta$ varies from $-\pi$ to $\pi$, the azimuthal angle $\varphi$ varies from 0 to $2\pi$, and $a$ is a scale parameter.

After some algebra one can get the denominator in Eq. (8) $1 - r^2 + 2iz = -(2/\tau)exp(-i\beta)$ and, therefore, the spinor components are $Z_1 = -\exp(-i\beta)/cosh(\eta)$, $Z_2 = tanh(\eta)exp(i\varphi)$, and $w(\mathbf{r}) = -sinh(\eta)exp[i(\varphi + \beta)]$. Such function $w(\mathbf{r})$ describes the hopfion with the azimuthal and poloidal vorticities, $n = 1$ and $m = 1$, respectively. The $z$-component of the magnetization is $m_z(\eta) = 1 - 2tanh^2(\eta)$ and varies from +1 to -1 for the interval $\eta \in [0,\infty)$. $m$, $n$ are integer numbers (winding numbers or vorticities in the poloidal and azimuthal directions), $m = \pm 1, \pm 2, ...$, $n = \pm 1, \pm 2, ...$, $a$ is the hopfion radius. The vorticities $m$ and $n$ are defined by the expression $m_x + im_y \propto exp[i(n\varphi + m\beta)]$ [8, 10, 11]. For arbitrary integer vorticities $m$, $n$ we can write for the azimuthally symmetric hopfion [12, 13]

$$w(\mathbf{r}, m, n) = \frac{[Z_2(\mathbf{r})]^n}{[Z_1(\mathbf{r})]^m}, \qquad (10)$$

where the spinor components $Z_2(\mathbf{r})$, $Z_1(\mathbf{r})$ are defined above.

The function $w(\mathbf{r}, m, n)$ in the toroidal coordinates $\mathbf{r}(\eta, \beta, \varphi)$ has the form

$$w(\mathbf{r}, m, n) = cosh^m(\eta)tanh^n(\eta)e^{i(m\beta+n\varphi)}. \qquad (11)$$

Therefore, the z-component of the hopfion magnetization defined by the relation $m_z = p(1 - |w|^2)/(1 + |w|^2)$ is

$$m_z(\eta, m, n) = p \frac{1 - \cosh^{2m}(\eta) \tanh^{2n}(\eta)}{1 + \cosh^{2m}(\eta) \tanh^{2n}(\eta)}, \tag{12}$$

where $p = m_z(\eta = 0)$ is the hopfion polarity, $p = \pm 1$.

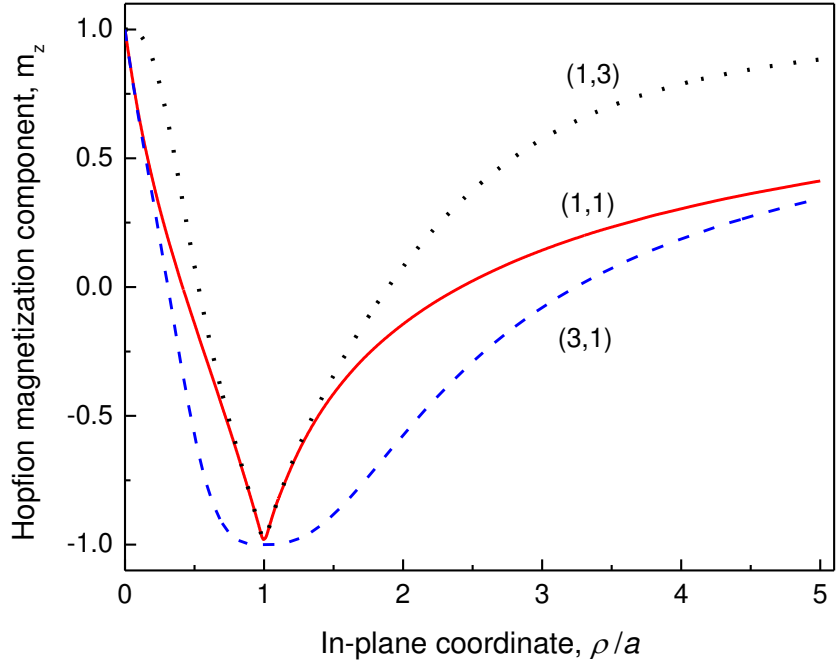

**Fig. 1. The hopfion magnetization profiles**, $m_z(\rho)$ in the plane z=0. The red solid line corresponds to $(m, n) = (1,1)$, the black dotted line - $(m, n) = (1,3)$, and the dashed blue line - $(m, n) = (3,1)$. The hopfion polarity $p = +1$.

The hopfion magnetization profiles, $m_z(\rho)$, calculated by using Eq. (12) for different values of the vorticities $(m, n)$ are shown in Fig. 1.

Other magnetization components in the toroidal coordinates can be found from the expression $m_x(\mathbf{r}) + im_y(\mathbf{r}) = \sqrt{1 - m_z^2(\eta)} exp[i(n\varphi + m\beta)]$. The particular values of the out-of-plane hopfion magnetization $m_z(0, m, n) = p$ (at the hopfion center) and $m_z(\infty, m, n) = -p$ (at the hopfion radius $\rho = a$, $z = 0$) are in agreement with ones defined in the paper [8]. For the particular case $n = 1$ and $m = 1$, the function $m_z(\eta, m, n)$ is reduced to the well-known form $m_z(\eta, 1,1) = p(1 - 2tanh^2(\eta))$ used by Hietarinta et al. [9]. The iso-surfaces of the constant magnetization component $m_z$ (or constant toroidal parameter $\eta$) are non-intersecting tori described by the equation $z^2 + (\rho - acoth(\eta))^2 = a^2/sinh^2(\eta)$. The torus centers are located at the rings $\rho = acoth(\eta)$ in the $xOy$-plane and the torus radii are equal to $a/sinh(\eta)$. The magnetization component is $m_z = +p$ in the hopfion center at $\mathbf{r} = 0$ and $m_z = -p$ at the ring $\rho = a$ in the limit $\eta \to \infty$ (see Fig. 1) that justifies using the scale parameter $a$ as the hopfion radius. The magnetic hopfion radius is analogues to the domain wall width in ferromagnets.

Let us calculate the emergent field vector potential, emergent magnetic field and the Hopf index of the magnetic hopfion represented via the complex function $w$ given by Eq. (10). Substitution the Hopf map (5) to the emergent field tensor (1) after some algebra using properties of the Pauli matrices, we get the emergent field tensor as a function of the spinor components

$$F_{\mu\nu}(Z) = -2i[\partial_\mu Z^+ \partial_\nu Z - \partial_\nu Z^+ \partial_\mu Z]. \tag{13}$$

The emergent field vector potential, which corresponds to the field tensor $F_{\mu\nu}(Z)$ is defined as

$$A_\mu(Z) = -i[Z^+\partial_\mu Z - \partial_\mu Z^+ Z]. \qquad (14)$$

Assuming using different orthogonal coordinate systems it is convenient to rewrite the expression (14) in the invariant form $\boldsymbol{A}(Z) = 2Im(Z^+\boldsymbol{\nabla}Z)$.

The Hopf index is the homotopy invariant of the mapping $S_3 \to S_2$. The emergent magnetic field does not depend on the gauge of the vector potential. However, the Hopf index is sensitive to a gauge choice of the vector potential $\boldsymbol{A}$. There is a definite, natural, gauge for which the non-zero Hopf index (4) can be calculated. To secure gauge invariance of the integral (4) calculated in the natural gauge (14), the emergent magnetic field $\boldsymbol{B}(\boldsymbol{r})$ has to satisfy the condition $\boldsymbol{\nabla} \cdot \boldsymbol{B} = 0$ (no emergent magnetic monopoles) and to vanish sufficiently fast at $|\boldsymbol{r}| \to \infty$. The first property is a consequence of the emergent field definition as $\boldsymbol{B} = \boldsymbol{\nabla} \times \boldsymbol{A}$. The second condition is valid for the localized magnetic solitons in infinite sample. Therefore, strictly speaking, the Hopf index defined by Eq. (4) can be gauge invariant only for infinite samples.

Using Eq. (14) one can calculate the vector potential components in any orthogonal coordinate system. The simplest calculations are in the toroidal coordinates $\boldsymbol{r}(\eta, \beta, \varphi)$. The normalized spinor components are

$$\tilde{Z}_1(\boldsymbol{r}) = \frac{1}{\sqrt{N(\eta)}} \frac{(-1)^m}{cosh^m(\eta)} e^{-im\beta}, \quad \tilde{Z}_2(\boldsymbol{r}) = \frac{1}{\sqrt{N(\eta)}} tanh^n(\eta) e^{in\varphi}, \quad (15)$$

where the normalization factor is $N(\eta) = 1/cosh^{2m}(\eta) + tanh^{2n}(\eta)$.

The vector potential components are

$$A_\eta = 0, \quad A_\beta = -\frac{2m\tau}{aN(\eta)cosh^{2m}(\eta)}, \quad A_\varphi = \frac{2n\tau tanh^{2n}(\eta)}{aN(\eta)sinh(\eta)}. \quad (16)$$

We note that the spinor components (15), the hopfion magnetization (12), and vector potential toroidal components (16) are essentially different from ones calculated in Ref. [10], where the particular form of the Lagrangian $\mathcal{L} = \pm(F_{\mu\nu}^2)^{3/4}$ was chosen. The spinor components, $Z_1, Z_2$, and the magnetization, $m_z(\eta)$, calculated in Ref. [10] depend on the ratio of the hopfion vorticities, $m/n$. In the present article we did not assume any particular form of the Lagrangian and calculated some consequences of the definition of the Hopf mapping by Eq. (5). The components (16) can be written immediately via the hopfion magnetization $m_z(\eta)$ defined by Eq. (12). Namely, we get $A_\eta = 0, A_\beta = -m\tau(m_z + 1)/a, A_\varphi = -n\tau(m_z - 1)/a sinh\eta$. Such components after using the correction multiplier (-2) coincide with ones obtained by Gladikowski et al. [8]. We call this choice of the vector potential **A** components as ¨natural¨ gauge. The corresponding emergent magnetic field components is easy to calculate using the equation $\boldsymbol{B} = \nabla \times \boldsymbol{A}$:

$$B_\eta = 0, \quad B_\beta = \frac{n\tau^2}{a^2 sinh(\eta)} \frac{\partial m_z}{\partial \eta}, \quad , B_\varphi = -\frac{m\tau^2}{a^2} \frac{\partial m_z}{\partial \eta}. \tag{17}$$

To clarify sense of the natural gauge and calculate the Hopf index of a hopfion magnetization texture we use the Hopf index representation via a volume integral (4). Then, using the volume element in the toroidal coordinates $dV = a^3 sinh(\eta) d\eta d\beta d\varphi / \tau^3$ and conducting the volume integration assuming infinite media we get the expression $Q_H = mn[m_z(0) - m_z(\infty)]/2$. According to Eq. (12) $m_z(0) = p$, $m_z(\infty) = -p$, and, therefore, the Hopf index $Q_H = mnp$ of the toroidal magnetic hopfion defined by the Hopf mapping (5) is integer for an infinite sample. The Hopf index (3D topological charge) is proportional to the product of the poloidal $m$ and azimuthal $n$ vorticities and does not depend on the details of the hopfion magnetization profile $m_z(\eta)$. However, the toroidal coordinates are inconvenient to describe any finite samples. The typical example of the finite sample, where the toroidal magnetic hopfion can be stabilized, is a cylindrical ferromagnetic dot. Therefore, we re-write Eq. (16), (17) in the cylindrical coordinates $(\rho, \varphi, z)$. Using the cylindrical coordinates, it is possible to interpret the Hopf index as a pure geometrical parameter.

We use the angular parameterization for the magnetization $\boldsymbol{m}(\boldsymbol{r})$ components via spherical angles $\Theta, \Phi$: $m_z = cos\Theta$, $m_x + im_y = sin\Theta exp(i\Phi)$, and the cylindrical coordinates $\boldsymbol{r}(\rho, \varphi, z)$ for the radius-vector $\boldsymbol{r}$, and the cylindrical components of the vector potential $\boldsymbol{A}$. The magnetization spherical angles are functions of the radius-vector, $\Theta = \Theta(\boldsymbol{r})$, $\Phi = \Phi(\boldsymbol{r})$. Following the theory of 2D magnetic solitons (vortices and skyrmions) [1], it is naturally to choose the hopfion magnetization spherical angles in axially symmetric form [7], $\Theta(\boldsymbol{r}) = \Theta(\rho, z)$, $\Phi(\boldsymbol{r}) = n\varphi + \gamma(\rho, z)$. The variable hopfion helicity $\gamma(\rho, z)$ is of principal importance to secure non-zero Hopf index and gyrovector of the 3D magnetization textures, such as a toroidal hopfion [13] or Bloch point hopfion [29], localized and non-localized magnetic topological solitons, respectively.

It follows from Eqs. (16), (17) that the cylindrical components of the vector potential $A$ and emergent magnetic field $B$ can be written as

$$A_\rho = -(1+m_z)\frac{\partial \gamma}{\partial \rho}, \quad A_\varphi = (1-m_z)\frac{n}{\rho}, \quad A_z = -(1+m_z)\frac{\partial \gamma}{\partial z} \qquad (18)$$

$$B_\rho = \frac{n}{\rho}\frac{\partial m_z}{\partial z}, \quad B_\varphi = \frac{\partial m_z}{\partial \rho}\frac{\partial \gamma}{\partial z} - \frac{\partial \gamma}{\partial \rho}\frac{\partial m_z}{\partial z}, \quad B_z = -\frac{n}{\rho}\frac{\partial m_z}{\partial \rho} \qquad (19)$$

The details on the calculations of the emergent magnetic field components for the particular case (*m*=1, *n*=1) and their connection with the hopfion gyrovector are given in Ref. [35]. For such particular case the simple relation $B(r) = [4a/(a^2 + r^2)]A(r)$ holds. There is the relation of the emergent field vector potential $A$ defined by Eq. (18) and the gauge vector potential $A^e = (1 - cos\Theta)\nabla\Phi$ defined as a component of the non-abelian gauge vector potential (forming a covariant derivative with respect to the spatial coordinates) along the local magnetization $m(r)$ direction in Ref. [32]:

$$\boldsymbol{A} = -2\nabla\gamma + \boldsymbol{A}^e \qquad (20)$$

The emergent magnetic field $\boldsymbol{B}$ (3) can be explicitly written via the magnetization spherical angles $\Theta$, $\Phi$ in the compact form $\boldsymbol{B} = \sin(\Theta)\nabla\Theta \times \nabla\Phi$. Therefore, in the $(\gamma, \Theta, \Phi)$ - representation the Hopf index (4) density $\boldsymbol{A} \cdot \boldsymbol{B} = -2\sin\Theta\nabla\gamma \cdot [\nabla\Theta \times \nabla\Phi]$ is the triple dot product as the vector potential $\boldsymbol{A}^e$ does not contribute to it. The Hopf index can be represented as an integral over the angles $(\gamma, \Theta, \Phi)$

$$Q_H = -\frac{2}{(4\pi)^2} \int d\gamma \int d\Theta \sin\Theta \int d\Phi. \tag{21}$$

Using the relations $\gamma = m\beta$ and $\partial\Phi/\partial\varphi = n$, the integration in Eq. (21) yields the integer Hopf index $Q_H = mnp$. Eq. (21) can serve as alternative definition of the Hopf index in comparison to the widely accepted its definition via the linking number of the preimages $\boldsymbol{r}(\boldsymbol{m}_1)$ and $\boldsymbol{r}(\boldsymbol{m}_2)$ in the 3D coordinate space of two points $\boldsymbol{m}_1$ and $\boldsymbol{m}_2$ of the unit sphere $S_2(\boldsymbol{m})$ [13-15, 26]. We also can conduct integration in Eq. (21) using the cylindrical coordinates $(\rho, \varphi, z)$. The Jacobian for the transformation from the $(\gamma, \Theta, \Phi)$ coordinates to the cylindrical coordinates is $J(\gamma, \Theta, \Phi/\rho, \varphi, z) = \rho^{-1} \partial\Phi/\partial\varphi [\partial\gamma/\partial z \, \partial\Theta/\partial\rho - \partial\Theta/\partial z \, \partial\gamma/\partial\rho]$. Therefore, the Hopf index is

$$Q_H = -\frac{2}{(4\pi)^2} \int dV_\rho \sin\Theta J = \frac{n}{4\pi} \int d\rho \int dz \sin\Theta \left[\frac{\partial\gamma}{\partial\rho}\frac{\partial\Theta}{\partial z} - \frac{\partial\gamma}{\partial z}\frac{\partial\Theta}{\partial\rho}\right]. \tag{22}$$

It is evident that the expression (22) coincides with the Hopf index calculated via the cylindrical components of **A** and **B** given explicitly by Eqs. (18), (19). According to Eqs. (16), (17) the Hopf charge density $\mathbf{A} \cdot \mathbf{B}$ in the toroidal coordinates is

$$\mathbf{A} \cdot \mathbf{B} = -2mn \frac{\tau^3}{a^3 sinh(\eta)} \frac{\partial m_z}{\partial \eta}$$

(23)

Accounting $dV = a^3 sinh(\eta) d\eta d\beta d\varphi / \tau^3$ and $\int d\eta \frac{\partial m_z}{\partial \eta} = \int dm_z$ the calculated Hopf index is equal to

$$Q_H = -\frac{2mn}{(4\pi)^2} \int dm_z \int d\beta \int d\varphi = -\frac{2mn}{(4\pi)^2} \int dV_\rho J_\eta, \tag{24}$$

where $J_\eta(m_z, \beta, \varphi/\rho, \varphi, z) = \rho^{-1}[\partial \beta / \partial \rho \, \partial m_z / \partial z - \partial m_z / \partial \rho \, \partial \beta / \partial z]$ is the Jacobian of the transformation from $(m_z, \beta, \varphi)$ coordinates to the cylindrical coordinates $(\rho, \varphi, z)$. Accounting the relation $m_z = m_z(\eta)$ these coordinates are closely related to the toroidal coordinates $(\eta, \beta, \varphi)$. The Hopf charge density in the cylindrical coordinates is determined by the Jacobians $J(\gamma, \Theta, \Phi/\rho, \varphi, z)$ or $J_\eta(m_z, \beta, \varphi/\rho, \varphi, z)$. The same is valid for any coordinate system: the Hopf charge density is determined by the Jacobian of transformation from the $(m_z, \beta, \varphi)$ coordinates to new orthogonal coordinates. The Hopf charge density $-2mn/(4\pi)^2$ is constant in the $(m_z, \beta, \varphi)$ coordinate representation. Similar identifies exist also for 2D topological charge (skyrmion number), where the magnetization $m(r)$ is function only of in-plane coordinates $r(x,y)$ or $r(\rho,\varphi)$. The skyrmion number $S$ is defined by the equation, which is analogous to Eq. (21):

$$S = \frac{1}{4\pi} \int d\Theta d\Phi \sin\Theta = \frac{1}{4\pi} \int dS_\rho \sin\Theta J, \qquad (25)$$

where the surface $S_\rho$ is the plane $z = const$ and the Jacobian $J(\Theta,\Phi/\rho,\varphi) = n\rho^{-1} \partial\Theta/\partial\rho$ corresponds to the transition from $(\Theta,\Phi)$ angles to the cylindrical $(\rho,\varphi)$ coordinates.

Let us use the spinor parametrization $Z_1 = cos(\Theta/2)exp(i\alpha)$, $Z_2 = sin(\Theta/2)exp(i(\Phi+\alpha))$ following Ref. [13] The complex function $w = Z_2/Z_1$ then has a well-known form $w = tan(\Theta/2)exp(i\Phi)$ and the magnetization components defined by Eq. (5) are $m_x + im_y = sin(\Theta)exp(i\Phi)$, $m_z = cos(\Theta)$. Neither the magnetization nor the emergent magnetic field $B$ depend on the gauge angle $\alpha(r)$. However, the emergent field vector potential $A$ does depend on $\alpha(r)$. Substituting $Z_1$, $Z_2$ to Eq. (14) one can get the expression $A = 2\nabla(\alpha+\gamma) + A_n$, where $A_n$ is the vector potential of the toroidal hopfion in the natural gauge obtained by Gladikowski et al. [8] in the toroidal coordinates (16) and by Wang et al. [24] in the cylindrical coordinates (18). Therefore, to keep the natural gauge and the integer Hopf index $Q_H = mnp$, we have to choose the gauge angle $\alpha(r) = -\gamma(\rho,z) + const$. The helicity has the form $\gamma(\rho,z) = m\beta(\rho,z)$, where the integer number $m$ is the poloidal vorticity, and $\beta$ is the poloidal angle in the toroidal coordinates $(\eta,\beta,\varphi)$ [34]. The function $\beta(\rho,z)$ has a pure geometrical origin related to transformation from the cylindrical to the toroidal coordinates. The explicit form of the functions $\eta(\rho,z)$ and $\beta(\rho,z)$ is given by the expressions $\eta(\rho,z) = atanh(2a\rho/(\rho^2 + z^2 + a^2))$, $\beta(\rho,z) = atan(2az/(\rho^2 + z^2 - a^2))$.

The authors of Refs. [36, 37] for the numerical calculations of the Hopf index used the emergent field vector potential gauge $A_z = 0$. According to Eq. (18) such choice of the vector potential is possible only if the hopfion magnetization component $m_z = -1$. The magnetization component $m_z$ is a function of the coordinates represented by Eq. (12) and can reach such extreme value only in the plane $z = 0$ at the polar radius vector $\rho = a$, see Fig. 1. Therefore, the condition $m_z = -1$ is unphysical. Nevertheless, the gauge $A_z = 0$ is formally possible to use because it leads to the correct expression for the Hopf index (4) of the infinite ferromagnetic media in the cylindrical coordinates given by Eq. (22). The gauge $A_z = 0$ essentially simplifies the numerical calculations of the Hopf index because there is only one nonzero component, $A_\varphi$, of the vector potential $\boldsymbol{A} = (A_\rho, A_\varphi, A_z) = (0, 2n/\rho, 0)$ in this gauge.

## 3. Conclusions

We considered the toroidal magnetic hopfions with the integer Hopf index $Q_H = mn$ and the arbitrary poloidal and azimuthal vorticities $m$, $n$ calculated in the toroidal and cylindrical coordinates. The calculation method is based on the Hopf mapping definition and the concept of the emergent magnetic field which is expressed via spatial derivatives of the magnetization field $\boldsymbol{m}(\boldsymbol{r})$. The hopfion magnetization is found explicitly for the arbitrary Hopf index. It is shown that the Hopf charge density can be represented as a Jacobian of the transformation from the toroidal to the cylindrical coordinates. The calculated components of the hopfion emergent magnetic field and vector potential can be used, in particular, for calculations of the topological Hall effect and skyrmion Hall effect of the toroidal magnetic hopfions, respectively.

The obtained above results are not specific for the unit magnetization field $\boldsymbol{m}(\boldsymbol{r})$. They are valid for any classical 3D field $\boldsymbol{n}(\boldsymbol{r})$ of the unit length $\boldsymbol{n}^2 = 1$ (order-parameter space is a unit sphere $S_2$) depending on the three spatial coordinates, which describes a physical system. For instance, in optics such field can be the Stocks vector field [17, 18], in the liquid crystals the role of the field plays the director (a molecular alignment) field [15, 36], etc.

**Acknowledgments:** The author acknowledges support by IKERBASQUE (the Basque Foundation for Science). The research was funded in part by the Spanish Ministry of Science and Innovation grant PID2019-108075RB-C33 /AEI/10.13039/501100011033 and the Norwegian Financial Mechanism 2014-2021 trough project UMO-2020/37/K/ST3/02450.